\documentclass[aps,showpacs,twocolumn,superscriptaddress]{revtex4}

\usepackage{graphicx}
\usepackage{dcolumn}

\begin{document}

\title{Comment on a recent attempt to explain the `GSI anomaly' \\ 
by initial-state e.m. spin-rotation coupling$^{\ast}$} 

\author{A.~Gal}\affiliation{Racah Institute of Physics, 
The Hebrew University, Jerusalem 91904, Israel} 

\begin{abstract} 

A recently proposed solution \cite{LPS13,LPS13a} of the `GSI anomaly' by spin 
precession of the decaying heavy ions in the magnetic field that controls 
their circular motion at the GSI storage ring is dubious: the uncertainty 
in the computed electron-capture decay-rate modulation frequency is at least 
of order $10^{7}$ Hz, by far exceeding the 1 Hz modulation frequency reported 
in the GSI experiment \cite{GSI08}. 

\end{abstract} 


\maketitle

Two-body electron-capture decay rates of $^{140}$Pr and $^{142}$Pm 
hydrogen-like heavy ions coasting at the GSI Experimental Storage Ring 
(ESR) were reported to be modulated with period $\sim$7~s \cite{GSI08}.
This modulation is known as the `GSI~anomaly' \cite{Kienlesummary}. 
Preliminary data exist also for $^{122}$I \cite{Kienle10}. 
It was concluded in Refs.~\cite{Faber10,Pavli10} that spin precession 
of the decaying ions in the ESR magnetic field $B\approx$1.2~T could not 
give rise to such modulation since the precession frequency associated with 
the electron is of order $|\mu_e| B\approx$3.3$\cdot$10$^{10}$~Hz and that 
associated with the nucleus is of order $\mu_p B\approx$2.5$\cdot$10$^{7}$~Hz, 
where $\mu_{e(p)}$ is the electron (proton) magnetic moment, both these 
frequencies exceeding by many orders of magnitude the reported modulation 
frequency $P_{\rm GSI}\sim 1$~Hz. 
We note that electron-capture decay rates of implanted neutral atoms, 
not exposed to external magnetic fields, do not show evidence for 
GSI-like modulation \cite{Vetter08,Faestermann09}.  

\begin{figure}[htb] 
\begin{center} 
\includegraphics[width=0.45\textwidth]{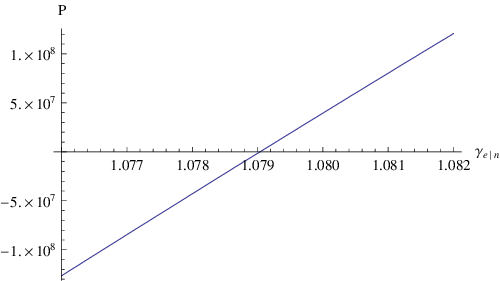} 
\caption{$P\equiv P_{\rm LPS}-P_{\rm GSI}$ (Hz units) in $^{140}$Pr 
as a function of the relativistic Lorentz factor $\gamma_{e|n}$. Here, 
$P_{\rm GSI}\sim 1$~Hz and $P_{\rm LPS}=a(\gamma_{e|n})\mu_e B+b(\gamma_{e|n},
g_{\rm nucl})\mu_p B$, $a$ and $b$ depending on $\gamma_{e|n}$ and $g_{\rm 
nucl}$. Note that a 0.1$\%$ variation of $\gamma_{e|n}$ induces $\approx$4$
\cdot$10$^{7}$ variation of $P$. Figure adapted from Ref.~\cite{LPS13}.} 
\label{fig:LPS} 
\end{center} 
\end{figure} 

Recently, however, Lambiase {\it et al.} \cite{LPS13,LPS13a} argued that 
a full treatment of spin-rotation coupling results in cancelations 
among components of the spin precession frequency $P_{\rm LPS}$, 
bringing $P_{\rm LPS}$ down to the level of $P_{\rm GSI}\sim 1$~Hz. 
Their $P_{\rm LPS}$ is responsible for splitting the two approximately 
$F$=$\frac{1}{2}$, $m$=$\pm\frac{1}{2}$ magnetic states of the rotating ion. 
Fig.~\ref{fig:LPS} adapted from Ref.~\cite{LPS13} shows the dependence 
of $P\equiv P_{\rm LPS}-P_{\rm GSI}$ on the relativistic Lorentz 
factor $\gamma_{e|n}$ in $^{140}$Pr. To sustain a zero value for $P$, 
$(\gamma_{e|n}-1)$ must be known accurately to at least {\it three} digits 
in order to trust the reduction of the $\mu_e$ term in $P_{\rm LPS}$ 
(see caption) from order $10^{10}$~Hz down to order $10^{7}$~Hz at 
which it might get canceled with the $\mu_p$ term; and moreover, the nuclear 
gyromagnetic ratio $g_{\rm nucl}$ must be known accurately to at least 
{\it seven} digits so as to reduce the order of magnitude of $P_{\rm LPS}$ 
from $10^{7}$~Hz further down to $P_{\rm GSI}\sim 1$~Hz. In practice, 
$(\gamma_{e|n}-1)$ is estimated semiclassically in Ref.~\cite{LPS13} to 
a single digit only, and $g_{\rm nucl}$ is either experimentally unknown 
or known to a single digit only: 
$g_{\rm nucl}(^{122}$I)=0.94$\pm$0.03~\cite{Green86}. Similar arguments 
were given by Faestermann~\cite{Faestermann09a} in response to earlier 
attempts by Lambiasse {\it et al.} to publish this kind of 
wishful thinking. 

In summary, atomic and nuclear magnetic moments are much too large to resolve 
the `GSI~anomaly' as attempted by Lambiasse {\it et al.} \cite{LPS13,LPS13a}. 
Magnetic moments of order $10^{-11}\mu_B$, many orders of magnitude below 
those provided by atoms and nuclei but less than one order of magnitude 
below the laboratory upper limit for neutrinos, are required to explain the 
`GSI anomaly' by spin precession~\cite{Gal10}.
\\ 

$^{\ast}${\it In memoriam} 
Paul Kienle (1931-2013) who pursued the `GSI~anomaly' with boundless 
imagination~\cite{GSI08,Kienlesummary,Kienle10,Faber10}. 

\end{document}